\documentclass{aastex631}
\pdfoutput=1
\usepackage[T1]{fontenc}
\usepackage{CJK}
\usepackage{amsmath}

\shorttitle{Detectability of stellar CMEs}
\shortauthors{Yang et al.}

\begin{document}

\title{Is it possible to detect coronal mass ejections on solar-type stars through extreme-ultraviolet spectral observations?}

\correspondingauthor{Hui Tian}
\email{huitian@pku.edu.cn}

\begin{CJK*}{UTF8}{gbsn}
\author[0000-0002-4973-0018]{Zihao Yang（杨子浩）}
\affil{School of Earth and Space Sciences, Peking University, Beijing 100871, People's Republic of China}
\affil{Deep Space Exploration Laboratory, Hefei, 230088, People's Republic of China}

\author[0000-0002-1369-1758]{Hui Tian（田晖）}
\affiliation{School of Earth and Space Sciences, Peking University, Beijing 100871, People's Republic of China}

\author[0000-0003-3908-1330]{Yingjie Zhu（朱英杰）}
\affiliation{Department of Climate and Space Sciences and Engineering, University of Michigan, Ann Arbor, MI 48109, USA}

\author[0000-0002-7421-4701]{Yu Xu（徐昱）}
\affiliation{School of Earth and Space Sciences, Peking University, Beijing 100871, People's Republic of China}

\author[0000-0002-1943-8526]{Linyi Chen（陈霖谊）}
\affiliation{School of Earth and Space Sciences, Peking University, Beijing 100871, People's Republic of China}

\author[0000-0001-5657-7587]{Zheng Sun（孙争）}
\affiliation{School of Earth and Space Sciences, Peking University, Beijing 100871, People's Republic of China}

\begin{abstract}
Stellar coronal mass ejections (CMEs) from host stars are an important factor that affects the habitability of exoplanets. Although their solar counterparts have been well observed for decades, it is still very difficult to find solid evidence of stellar CMEs. Using the spectral line profile asymmetry caused by the Doppler shift of erupting plasma, several stellar CME candidates have been identified from spectral lines formed at chromospheric or transition region temperatures of the stars. However, a successful detection of stellar CME signals based on the profile asymmetries of coronal lines is still lacking. It is unclear whether we can detect such signals. Here we construct an analytical model for CMEs on solar-type stars, and derive an expression of stellar extreme-ultraviolet (EUV) line profiles during CME eruptions. For different instrumental parameters, exposure times, CME propagation directions and stellar activity levels, we synthesized the corresponding line profiles of Fe \sc{ix}\rm{} 171.07 \AA\ and Fe \sc{xv}\rm{} 284.16 \AA. Further investigations provide constraints on the instrumental requirements for successful detection and characterization of stellar CMEs. Our results show that it is possible to detect stellar CME signals and infer their velocities based on spectral profile asymmetries using an EUV spectrometer with a moderate spectral resolution and signal-to-noise ratio. Our work provides important references for the design of future EUV spectrometers for stellar CME detection and the development of observation strategies.

\end{abstract}


\section{Introduction} \label{sec:intro}

Coronal mass ejections (CMEs) are large-scale solar eruptions associated with massive amounts of plasma expulsion and release of enormous energy. In the solar system, CMEs play an important part in shaping the interplanetary space environment, driving the variations in space weather around Earth and other planets \citep[e.g.,][]{2016GSL.....3....8G,2022SpWea..2003215P}. CMEs also occur on other stars (stellar CMEs). Stellar CMEs are one of the primary factors determining the habitability of exoplanets through processes such as atmospheric erosion and other space weather effects \citep[e.g.,][]{2007AsBio...7..167K,2007AsBio...7..185L,2020IJAsB..19..136A}. Additionally, they may have significant impacts on stellar evolution through stellar mass loss and angular momentum loss \citep[e.g.,][]{2012ApJ...760....9A,2017MNRAS.472..876O}.

Solar CMEs have been well observed through multiple approaches including imaging observations from white-light coronagraphs \citep[e.g.,][]{2011JGRA..116.4104W,2012ApJ...751...18F,2023ApJ...952L..22S} and extreme-ultraviolet (EUV) imagers \citep[e.g.,][]{2014ApJ...780...28C,2018ApJ...868..107V}. A few successful observations have also been made through solar spectral observations at different wavelengths \citep[e.g.,][]{2010ApJ...711...75L,2012ApJ...748..106T,2013SoPh..288..637T,2013JGRA..118..967G,2022ApJ...931...76X,2023ApJ...953...68L}. 

However, it is still difficult to detect their counterparts on other stars. Only a few observational attempts have been made on the detection of stellar CMEs using different methods. One of the approaches is the coronal dimming method. During solar CMEs, the ejected material (mass loss) may result in a sudden decrease in the EUV and X-ray emission, which is the coronal dimming phenomenon \citep[e.g.,][]{2012ApJ...748..106T,2014ApJ...789...61M}. Coronal dimming can be a signature of CME eruptions, and is useful in inferring the mass and speed of CMEs \citep{2012ApJ...748..106T,2016ApJ...830...20M}. Using EUV and X-ray observations of late-type stars, \cite{2021NatAs...5..697V} identified different stellar CME candidates by extrapolating solar coronal dimming techniques to stellar cases. Another possible approach to detecting stellar CME candidates is through X-ray absorption. \cite{2017ApJ...850..191M} identified a ``monster" stellar CME candidate from the continuous dropping down of X-ray emission, which was interpreted as the obscuration by stellar CMEs or prominences. Radio observations could also shine a light on the discovery of stellar CMEs. During solar CMEs, it has been widely believed that the fast shocks formed in front of CMEs can produce type-II radio bursts. It is also possible to observe type-II radio bursts from stellar observations. However, no strong evidence has been publicly reported so far, one reason could be related to the frequency of type-II radio bursts caused by some stellar CMEs being below the Earth's ionospheric cutoff frequency \citep[e.g.,][]{2020ApJ...895...47A}. Of all the different approaches, a promising method is through the detection of spectral profile asymmetry caused by erupting stellar CMEs from spectral observations. During solar CMEs, the propagation of a large amount of plasma will contribute to a secondary blue-shifted or red-shifted component of the spectral line profiles, leading to spectral profile asymmetry. Using the spatially resolved EUV spectral observations from Hinode/EIS \citep[][]{2007SoPh..243...19C}, \cite{2012ApJ...748..106T} found blue-wing enhanced asymmetric profiles from different EUV spectral lines during a mass ejection event. From Sun-as-a-star EUV spectral observations of SDO/EVE, \cite{2022ApJ...931...76X} and \cite{2023ApJ...953...68L} also found several examples of blue-wing enhancement in spectral profiles caused by CMEs. An analytical model was also used to investigate the detectability of solar CMEs from Sun-as-a-star EUV spectroscopic observations \citep{2022ApJS..260...36Y}. 

Using the spectral profile asymmetry technique, dozens of stellar CME candidates have been reported on various types of stars. Most of the candidates were discovered using lines formed at chromospheric or transition region temperatures. For example, \cite{1990A&A...238..249H} observed strong blue-wing enhancements in H-$\gamma$ and H-$\delta$ lines from M dwarf AD Leo, which were interpreted as the signal of a violent mass ejection event; using H-$\alpha$, H-$\beta$ and H-$\gamma$ spectroscopic observations for different targets, \cite{2019A&A...623A..49V} also reported possible stellar CME candidates; H-$\alpha$ line asymmetry observed from a solar-type star EK Dra indicates another possible stellar CME \citep{2021NatAs...6..241N}; blue-wing enhancements observed in H-$\alpha$ and Mg \sc{i}\rm{} triplet lines hint for a possible detection of stellar CMEs from G, K and M-type stars \citep{2022A&A...663A.140L}. Similar signals have also been observed from slightly hotter C \sc{iii}\rm{} lines in the far ultraviolet (FUV) passband \citep{2011A&A...536A..62L}. \cite{2019NatAs...3..742A} detected several stellar CME candidates from soft X-ray spectroscopic observations, but using spectral lines formed at transition region temperatures  for a giant star which is off the main-sequence. Using soft X-ray spectroscopic observations of the M-type flare star EV Lac, \cite{2022ApJ...933...92C} detected a blue shift and density decrease using spectral lines formed at transition region temperatures of the star, which is a possible evidence for a mass ejection event. However, none of the above observations were made with spectral lines formed in the stellar coronae of main-sequence stars. From the perspective of searching for habitable exoplanets, main-sequence stars are the most relevant type of host stars since they will stay in the main sequence for a very long time before burning out of the hydrogen, especially for the long lifetime solar-type and late-type stars \citep[e.g.,][]{2013ApJ...765..131K,2013AsBio..13..833R}. Stellar CMEs from G, K and M stars may have a large impact on the habitability of the exoplanets in their habitable zones \citep[e.g.,][]{2019LNP...955.....L}. It is important to consider the impacts of stellar CMEs when studying the habitability of exoplanets. Up to now, we still lack evidences from their coronal spectral observations of solar-type and late-type stars. One reason is related to suitable instruments. Coronal emission of G, K and M stars falls largely into the in EUV wavelength range, however, only the Extreme Ultraviolet Explorer spectrometer \citep[EUVE,][]{1985ApOpt..24.1737H} used to obtain EUV spectra of only $\sim$15 main-sequence stars. Besides the requirement for suitable instruments, the strong absorption of the interstellar medium makes it difficult to observe stellar spectra in the longer wavelength of EUV \citep{1994AJ....107.2108R}. In consideration of both factors, to obtain observational evidences of stellar CMEs using EUV spectra, a new EUV spectrometer is needed, with its passpand limited to the shorter wavelength range. To design such an instrument, it is necessary to investigate the constraints on instrumental configurations for successful detection of stellar CME signals, as well as precise measurements of their velocities.

In this work, we extend our previous analytical investigation based on Sun-as-a-star spectroscopy from the detection of solar CMEs \citep[][hereafter Paper I]{2022ApJS..260...36Y} to detection of stellar CMEs on solar-type stars. We modify the CME model to make it suitable for typical solar-type stars and derive a new integral expression for full disk-integrated spectral line profiles during stellar CME eruptions on such stars. Different spectral line profiles are synthesized under different CME propagation directions, exposure times and stellar activity levels. Using a similar quantitative criterion, we constrain the instrumental configurations for successful detection of stellar CME velocities. The impact of interstellar absorption is also discussed. Our work answers the question of how to detect stellar CME signals from EUV spectral observations, and provides important information on the instrumental requirements for future stellar EUV spectrometers. It also provides insights into the observational strategy to detect stellar CMEs on solar-type stars using the spectral profile asymmetry method.

\section{Methodology}

In this section, we describe the methodology to synthesize spectral line profiles during stellar CMEs. The methodology consists of the following parts: 1. Reference line profiles (Section \ref{sec:refline}): we use reference line profiles from solar observations as a baseline to calculate the spectral line profiles of stellar CMEs. 2. The models and assumptions of the stellar CME (Section \ref{sec:model}). 3. The calculation of the spectral line emission (Section \ref{sec:calcemission}), spectral line profiles (Section \ref{sec:calcprofile}) and the synthesis of full disk-integrated line profiles during stellar CME eruptions (Section \ref{sec:synthesis}).

\subsection{Reference line profiles}\label{sec:refline}
As we have no previous information on the EUV line profiles during CMEs of solar-type stars, we use the spectral observations from our nearest G-type star, the Sun, as our references.
We started with the same EUV spectral observation of an erupted loop from Hinode/EIS, which was previously used in \cite{2012ApJ...748..106T} and Paper I. The observation provides spectral line profiles in background active region (AR) and CME region for Fe \sc{xii}\rm{} 195.12 \AA\ and Fe \sc{xv}\rm{} 284.16 \AA. Previous observations of solar jets indicate that jets or mass ejection regions usually have densities several times higher than the surrounding background regions \citep[e.g.,][]{2012ApJ...748..106T}. From MHD modeling of CMEs, it is also conventionally assumed that CME regions have densities approximately three times higher than the static coronal regions \citep[e.g.,][]{2019ApJ...884L..13A}. We thus assumed that, for solar-like stars, the electron density in CME regions could be three times higher than that in the background active regions. Considering the EIS observations we used, the densities in CME region and background ARs are found to be remarkably similar \citep{2012ApJ...748..106T}. Therefore, we assumed a new CME density three times larger than that of the CME regions in the original EIS observation. As the EUV optically thin line intensity is proportional to the square of electron density ($n_e^2$), we could readily calculate the intensity and line profiles of the corresponding line using a threefold increase in electron density. Further details regarding this calculation will be described in Appendix \ref{app:appendixa}.

Figure \ref{fig: ref_profile} presents the reference line profiles of Fe \sc{ix}\rm{} 171.07 \AA\ and Fe \sc{xv}\rm{} 284.16 \AA\ at the onset of the eruption. Each line profile comprises two components: the CME component and the background AR component. The Fe \sc{xv}\rm{} 284.16 \AA\ line profile was calculated based on actual solar observations as mentioned earlier. However, instead of using the observed Fe \sc{xii}\rm{} 195.12 \AA, we have chosen Fe \sc{ix}\rm{} 171.07 \AA\, which was not observed by EIS, as our target line. This is because Fe \sc{xii}\rm{} 195.12 \AA\ is strongly affected by interstellar absorption (as demonstrated in the subsequent sections), thus making the cleaner and bright Fe \sc{ix}\rm{} 171.07 \AA\ more suitable for our analysis. To generate the corresponding AR line profile of Fe \sc{ix}\rm{} 171.07 \AA\ under the same condition, we used a fixed intensity ratio between Fe \sc{ix}\rm{} 171.07 \AA\ and Fe \sc{xii}\rm{} 195.12 \AA\ (observed in the aforementioned dataset) in active regions. The detailed calculation methodology will be described in Appendix \ref{app:appendixa}.

Past observations of Hinode/EIS have revealed that the ratios of typical EUV spectral lines in active regions and quiet Sun (QS) regions fluctuate in some ranges \citep[e.g.,][]{2008ApJS..176..511B}, but a fixed intensity ratio value between AR and QS within the reasonable ranges can be adopted as the typical intensity ratio (e.g., Paper I). In our calculations, we adopted a fixed value of 7 as the AR/QS ratio for Fe \sc{ix}\rm{} 171.07 \AA, and a value of 375 as the AR/QS ratio for Fe \sc{xv}\rm{} 284.16 \AA\ \citep[e.g.,][]{2007PASJ...59S.857Y,2022ApJS..260...36Y}. Fe \sc{xv}\rm{} 284.16 \AA\ is formed in hotter regions with its formation temperature of $\log_{10}T/K\sim 6.35$, while the cooler Fe \sc{ix}\rm{} 171.07 \AA\ has a formation temperature of $\log_{10}T/K\sim 5.9$. Therefore, the intensity of Fe XV 284.16 \AA\ will be much weaker in the relatively cooler quiet Sun regions, making its AR/QS ratio much larger than that of Fe IX 171.07 \AA. Based on the AR/QS ratios, we can also obtain the reference profiles in QS regions. The reference profiles in AR and QS are denoted as $\bar{I}_{\text{AR}}$ and $\bar{I}_{\text{QS}}$, respectively. Note that we only use the intensities and line widths of these reference line profiles.

\begin{figure}[htbp!]
\centering
\includegraphics[width=0.8\textwidth]{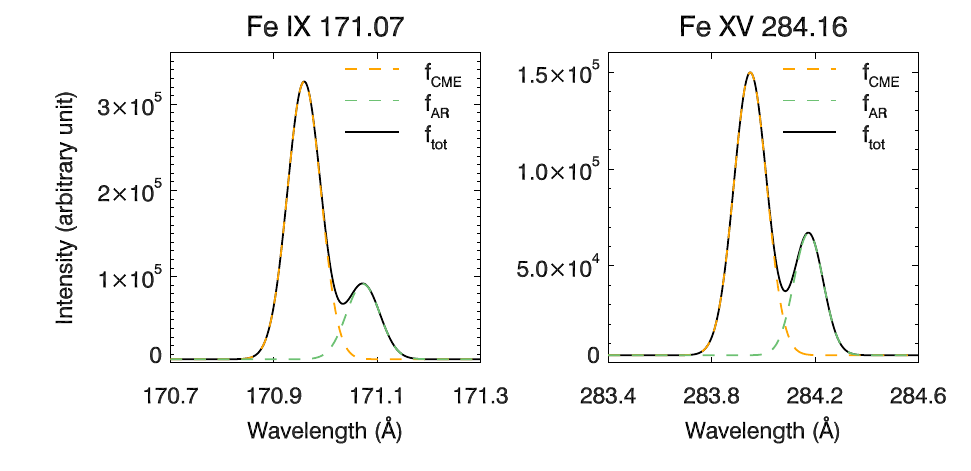}
\caption{The reference line profiles of Fe \sc{ix}\rm{} 171 \AA\ (left) and Fe \sc{xv}\rm{} 284 \AA\ (right) at the onset of the CME eruption. The orange dashed profiles are the spectral line profiles contributed by the CME, the green dashed profiles are emission from background active regions, and the black solid profiles are the total line profiles.}  
\label{fig: ref_profile}
\end{figure}

\subsection{Stellar CME model}\label{sec:model}
Our stellar CME model consists of two parts: the kinematic model, which characterizes the height and velocity evolution of the CME during its initial propagation, and the geometric model, which describes the shape and structure of the stellar CME. The two models are coupled together to form a comprehensive model describing the evolution of the CME.

\subsubsection{Kinematic Evolution Model of the Stellar CME}\label{sec:kinemodel}

To investigate the evolution of CME emissions during its propagation, we adopted a kinematic profile of CMEs from \cite{2003ApJ...588L..53G}. The acceleration rate of the CME is time-dependent, with an exponential increase followed by an exponential decrease:

\begin{equation}
    a(t)=\left[\frac{1}{a_r\exp{(t/t_r)}}+\frac{1}{a_d\exp{(-t/t_d)}}\right]^{-1}
\end{equation}
where $a_r=1$ m s$^{-2}$, $t_r=164$ s, $a_d=4024$ m s$^{-2}$, and $t_d=1127$ s.

The velocity of the CME is then 

\begin{equation}
    v(t)=v_0+\int_{0}^t a(t)\text{d}t
\end{equation}
where $v_0=36$ km s$^{-1}$. The height function is then a time integration of the velocity function. 

\begin{figure}[htbp!]
\centering
\includegraphics[width=0.8\textwidth]{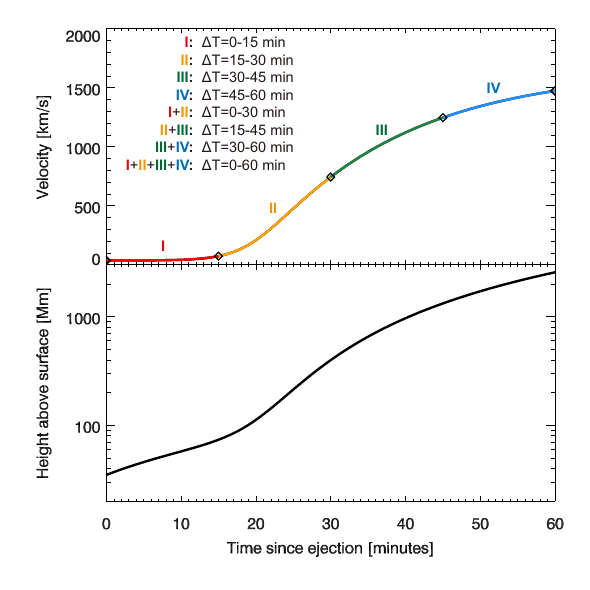}
\caption{The temporal evolution profiles of the velocity and height of the stellar CME. To investigate the impacts of different exposure times on the observed spectral line profiles, we selected several different integration time periods (15 min, 30 min and 60 min). The four solid sections correspond to four different time ranges with an integration time of 15 min. Other integration times are from different combinations of the four 15-min sections.}  
\label{fig: evo_profile}
\end{figure}

The evolution profiles of CME height and velocity are shown in Fig. \ref{fig: evo_profile}. This CME exhibits a comparatively slow propagation pattern compared to the kinematic models in Paper I. As mentioned in Paper I, the velocity and density of the CME material vary during the propagation, leading to changes in the observed line profiles contributed by the CME material. We will provide more details on this in the following sections.

One important parameter in stellar observations is the exposure time. In Paper I, the effect of exposure time was disregarded due to the focus on solar observations. However, for stellar observations, longer exposure times are often required to acquire sufficient signal-to-noise ratio (S/N). As the stellar CME evolves during the exposure, the observed line profiles represent an integrated result from line profiles formed at different heights with varying CME velocities. Different exposure times, or integration times, can significantly impact the observed line profiles. To investigate this effect, we selected three different exposure times: 15 min, 30 min, and 60 min. For the exposure time of 15 min, we calculated the integrated line profiles within the time ranges of 0-15 min, 15-30 min, 30-45 min and 45-60 min. For the exposure time of 30 min, we performed integration in the time ranges of 0-30 min, 15-45 min and 30-60 min. An extended long exposure of 60 min is also adopted in the calculation of integrated line profiles.

\subsubsection{Geometric model of the stellar CME}\label{sec:geomodel}

Similar to Paper I, we adopted a simplified graduated cylindrical shell (GCS) model \citep{2006ApJ...652..763T} to represent the CME flux tube, and incorporated the assumptions of self-similar expansion, uniform density distribution and mass conservation during CME propagation. We refer to Paper I for a detailed description of the geometry. Figure \ref{fig:stellar CME model}(A) shows our geometric model with the CME flux tube represented as a cylindrical structure with a radius of $r$ in its symmetric plane. Figure \ref{fig:stellar CME model}(B) shows the CME symmetric plane on the equatorial plane, with the CME structure extending radially towards the observer (the vector pointing from the stellar center to the center of the CME torus segment is along the LOS). Figure \ref{fig:stellar CME model}(C) represents that the CME symmetric plane is still on the equatorial plane but the direction from the stellar center to the CME torus center is rotated by an angle of $\eta$ with respect to the LOS direction (the azimuth). In Figure \ref{fig:stellar CME model}(D), the CME symmetric plane is further rotated with an angle $\xi$ around the rotated Y axis that is perpendicular to the line connecting the stellar center to the sub-CME point. These two angles ($\eta,\xi$) are used to describe the propagation direction of the CME. A further explanation will be provided in the subsequent sections.

\subsection{Calculation of Spectral Line Emission}\label{sec:calcemission}

We mainly refer to Equations (3)-(9) in Paper I for the calculation of spectral line emissions. 

In Paper I, it is noted that the upper levels of the optically thin EUV lines in the selected waveband are primarily populated through the process of electron collisional excitation, consequently the emissivity of the EUV lines should in principle be proportional to the square of electron density, i.e., $i\propto n_e^2$. However, this relationship holds true only when the contribution function of the target line is solely dependent on temperature. In fact, the contribution function is also density-dependent, indicating that the sensitivity of the emissivity to $n_e^2$ varies with different $n_e$. 

The intensity of an optically thin EUV line can be expressed as follows \citep{2018LRSP...15....5D}
\begin{equation}\label{eq:gofnt}
        I(\lambda)=\int{G(T,n_e)\cdot n_en_H}\ \text{d}s\sim \int{G(T,n_e)\cdot n_e^2}\ \text{d}s
\end{equation}
where $I(\lambda)$ is the spectral intensity at wavelength $\lambda$, $G(T,n_e)$ is the contribution function, and d$s$ is the differential element of distance along the LOS.

The contribution function $G(T,n_e)$ can also be represented as 
\begin{equation}
    G=\frac{hc}{4\pi\lambda}\cdot Ab(Z)\cdot Ab(Z^{+r})\cdot\frac{N_j(Z^{+r})}{N(Z^{+r})}\cdot A\cdot\frac{1}{n_e}
\end{equation}
where $Ab(Z)$ denotes the element abundance, $Ab(Z^{+r})$ is the ion abundance, $A$ is the Einstein spontaneous emission coefficient, $N_j(Z^{+r})$ is the population of the upper level $j$, and $N(Z^{+r})$ is the number density of the ion $Z^{+r}$. $Ab(Z),\ Ab(Z^{+r})$ are nearly independent of the electron density, and $A$ is independent of the electron density \citep[e.g.,][]{2018LRSP...15....5D}. 

Here we denote the term $\dfrac{N_j(Z^{+r})}{N(Z^{+r})}\dfrac{A}{n_e}$ as $Q(n_e)$. To calculate the dependence of $Q(n_e)$ on electron density $n_e$, we use the function \it{emiss\_calc.pro}\rm{} in CHIANTI ver. 10 \citep{2021ApJ...909...38D}. By comparing the expression of $Q(n_e)$ with Eq. \ref{eq:gofnt}, we find that the results of function \it{emiss\_calc.pro}\rm{} is equivalent to the dependence of the contribution function on electron density. Here the temperature is set as the formation temperature of the corresponding spectral line. Therefore, we have 
\begin{equation}
    i=\frac{P}{4\pi}\propto G\cdot n_e^2\propto Q(n_e)\cdot n_e^2
\end{equation}

In addition to electron collisional excitation, the upper energy levels of the target lines can also be populated through photo-excitation, albeit to a very small extent. $Q(n_e)$ depends on $(n_e)$ (which is solely dependent on height according to Equation (3) in Paper I) and photo-excitation at different heights. Therefore we calculated $Q(n_e)$ as a function of height $h$ using CHIANTI database
\begin{equation}\label{eq:qfunc}
    i(h)\propto Q(h)\cdot n_e^2(h)
\end{equation}
This implies that the unit volume intensity $i$ is not simply proportional to $n_e^2$, but it varies differently with different electron density.

Combining Eq. \ref{eq:qfunc} with Equation (3) in Paper I, the expression of $i(h)$ as a function of height $h$ can be obtained
\begin{equation}\label{eq:iofqh}
    \frac{i(h)}{i(h_0)}=\frac{Q(h)}{Q(h_0)}\left(\frac{n_e(h)}{n_e(h_0)}\right)^2=\frac{Q(h)}{Q(h_0)}\left(\frac{R_s+h}{R_s+h_0}\right)^{-6}
\end{equation}
where $h_0$ is the initial height of EIS-observed eruption (in this paper $h_0\sim 34.7$ Mm). Here we introduced the impact of density dependence to the original Equation (13) in Paper I by considering the density dependence of contribution functions.

The variation of the unit continuum intensity $i_{\text{cont}}$ with height has been described in Equation (14) in Paper I.

\subsection{Calculation of Spectral Line Profiles}\label{sec:calcprofile}
The geometric configuration and expansion pattern of the CME described in Paper I remain unchanged in this study. As illustrated in Fig. \ref{fig:stellar CME model}, the CME is assumed to have a bulk velocity $\mathbf{v}_{\text{bulk}}$ and an expansion speed $\mathbf{v}_{\text{exp}}(d)$. For our analysis, we consider the flux tube to be composed of multiple concentric shells. Moving from the the center to the outermost layer, the radius of each shell ($d$) increases from 0 to $r$. The expansion speed is a function of $d$ and linearly increases from $d=0$ to $d=r$. Following the derivations in Section 2.4 of Paper I, we obtain the expression for the expansion speed:
\begin{equation}\label{eq:vexp}
    v_{\text{exp}}(d)=\frac{d}{R_s+h}v_{\text{bulk}}
\end{equation}
where $v_{\text{bulk}}$ is extracted from the velocity profile in Fig. \ref{fig: evo_profile}. The total velocity (vector) is $\mathbf{v}=\mathbf{v}_{\text{bulk}}+\mathbf{v}_{\text{exp}}(d)$. It is to be noted that $\mathbf{v}_{\text{bulk}}$ and $\mathbf{v}_{\text{exp}}(d)$ are generally not parallel.

\subsubsection{LOS speed}
\begin{figure}[htbp!]
\centering
\includegraphics[width=0.95\textwidth]{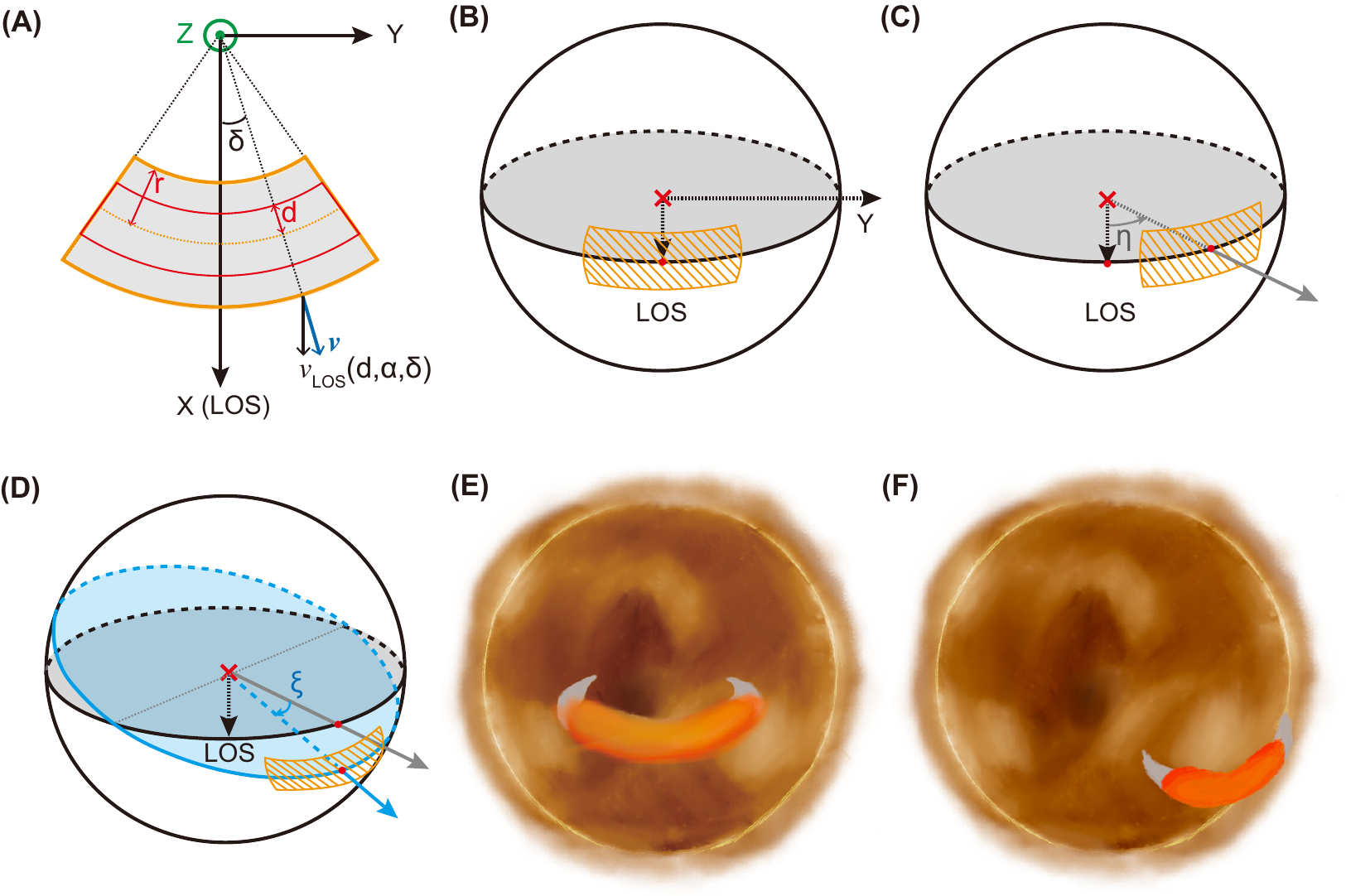}
\caption{The stellar CME geometric model. We refer to the partial donut-shape structure or curved cylinder from Figure 3 in Paper I as the CME flux tube. (A) The CME on its symmetric plane (the XY plane). The CME is a face-on eruption lying on the XY plane. X is the LOS direction. Here the CME structure is decomposed into multiple concentric shells. $r$ is the radius of the CME flux tube, and $d$ is the radius of the shell within the CME flux tube ranging from 0 to $r$. $\delta$ is the toroidal azimuthal angle on the toroidal cross section and $\alpha$ is the poloidal azimuthal angle on the poloidal cross section (see Figure 3 in Paper I). We refer to Paper I for details of LOS velocity and vector velocity at any positions of the CME structure. (B) The initial position of the CME. Here the orange shaded region is the radial projection of the CME onto the stellar surface. The sub-CME point, located at the center of the stellar disk, points in the direction from the stellar center to the observer. The CME symmetric plane aligns with the equatorial plane. (C) Similar to (B) but showing the position of the CME after a rotation of $\eta$. Here the CME torus is rotated with an angle $\eta$ within the equatorial plane (its symmetric plane), resulting in a rotation of the sub-CME point. Consequently, the sub-CME point now points at an angle of $\eta$ relative to the LOS. (D) Similar to (B)(C) but showing the position of the CME after a second rotation of $\xi$. The CME's symmetric plane undergoes rotation along the rotated Y axis that is perpendicular to the line connecting the stellar center to the sub-CME point. This rotation inclines the symmetric plane at an angle of $\xi$. (E) A 3D cartoon illustrating the stellar CME geometry at its initial position (corresponding to (B)). (F) Similar to (E) but for the stellar CME after both rotations (corresponding to (D)).}  \label{fig:stellar CME model}
\end{figure}

The LOS velocity is a key parameter to be determined from spectral observations. In this part, we adopt a similar approach as in Paper I to derive the LOS velocity. In Paper I, only Earth-facing CMEs were considered, meaning that the bulk velocity of the CME is purely along LOS direction (Figure \ref{fig:stellar CME model}(A)). Figure \ref{fig:stellar CME model}(B) illustrates the initial position of the radial projection of the CME, represented by the orange shaded region. The sub-CME point is located at the center of the stellar disk. In other words, the center of the CME torus traces a curve that resembles a segment of a circular orbit on the stellar equatorial plane. However, this does not always hold in real observations. The observed CMEs could propagate along different directions, and the propagation direction (defined as the angle between bulk velocity and LOS) greatly affects the derived LOS speed. As depicted in Fig. \ref{fig:stellar CME model}(C), we performed a rotation of the CME torus within the equatorial plane, adjusting its orientation by an angle of $\eta$. As a result, the CME's symmetric plane remains aligned with the equatorial plane, while the sub-CME point undergoes a rotation.  Subsequently, we proceeded to rotate the CME's symmetric plane (equatorial plane) with an inclination angle of $\xi$, along a direction perpendicular to the line connecting the stellar center to the sub-CME point. This configuration is illustrated in Figure \ref{fig:stellar CME model}(D). Following these two rotations, the position of the sub-CME point can be described in terms of the azimuth angle $\eta$ and the inclination angle $\xi$. In this work, to demonstrate the impact of different propagation directions, we choose three distinct propagation directions in our synthesis: $(\eta=0^{\circ},\ \xi=0^{\circ}),\ (\eta=30^{\circ},\ \xi=30^{\circ})$ and $(\eta=60^{\circ},\ \xi=60^{\circ})$. Figure \ref{fig:stellar CME model}(E) is a 3D illustration for the case of $(\eta=0^{\circ},\ \xi=0^{\circ})$, while Figure \ref{fig:stellar CME model}(F) is an illustration for tilted-propagating CME.  We do not aim to perform calculations on every possible propagation directions, but rather to illustrate the effect of varying propagation directions on the synthesized profiles. 

For face-on eruptions, the velocity vector can be derived following Figure 3(D) and Eq. (28) in Paper I, as well as from Fig. \ref{fig:stellar CME model}(A) as 
\begin{equation}\label{eq:vecvel}
    \mathbf{v}_0=(v_x,v_y,v_z)^{T}=
    \left[\begin{array}{c}
    v_{\text{bulk}}\cos\delta+v_{\text{exp}}(d)\cos\alpha\cos\delta\\
    v_{\text{bulk}}\sin\delta+v_{\text{exp}}(d)\cos\alpha\sin\delta\\
    v_{\text{exp}}(d)\sin\alpha
    \end{array}\right]
\end{equation}
Different propagation direction $(\eta,\xi)$ is equivalent to rotating the CME structure with Euler angles of $\eta$ and $\xi$. Figure \ref{fig:stellar CME model}(C) and (D) shows the rotation of the CME structure with the two rotation angles $\eta$ and $\xi$. Therefore, the velocity vector after rotation can be derived by multiplying the original velocity vector (Eq. \ref{eq:vecvel}) with the rotation matrix. In this case, the LOS velocity after the rotation is
\begin{equation}\label{eq:losv}
\begin{split}
    v_{\text{LOS}}(d,\alpha,\delta,\eta,\xi)&=R\cdot\mathbf{v}_0\cdot\mathbf{e}_x\\
    &=[v_{\text{bulk}}\cos\delta+v_{\text{exp}}(d)\cos\alpha\cos\delta]\cdot\cos\eta\cos\xi-[v_{\text{bulk}}\sin\delta+v_{\text{exp}}(d)\cos\alpha\sin\delta]\cdot\sin\eta\\&+v_{\text{exp}}(d)\cos\eta\sin\alpha\sin\xi
\end{split}   
\end{equation}
where $R$ is the rotation matrix. This equation represents the LOS speed at every part of the flux tube. Here $d\in[0,r]$ and $\delta\in[-\theta_0,\theta_0]$. the value of $\theta_0$ is determined by the CME area at a specific height (the initial height, see Equations (10)-(12) in Paper I). in this work, we consider the pairs of $\eta$ and $\xi$ as $(0^{\circ},0^{\circ}),\ (30^{\circ},30^{\circ})$ and $(60^{\circ},60^{\circ})$. Note that we do not include a third Euler angle to account for the roll angle (the rotation around the CME structure's own axis of symmetry) for simplicity of calculation. However, the readers can easily incorporate this angle into the rotation matrix $R$ and derive the expression for the LOS velocity.

\subsubsection{Line Profiles}\label{sec:lineprof}

Following the derivations in Sect. 2.4.2 and Figure 3(E) of Paper I, we can decompose the CME flux tube into numerous differential elements with the volume $\text{d}V$ (see Eq. (23) in Paper I). Each differential element has the flux of $\text{d}F=\text{d}V\cdot i/D^2$, where $i$ is the unit volume intensity at a given height $h$. The unit volume line intensity ($i$) and unit volume continuum ($i_{\text{cont}}$) in each differential volume are constant based on the uniform distribution assumption. Additionally, if we consider the line widths (exponential width or 1/e width $w$) to be constant for ARs, QS and CMEs as the CME propagates, we can describe the line profile formed within each differential element of the CME as
\begin{equation}
    \text{d}f(\lambda)=\frac{\text{d}F}{\sqrt{\pi}w}\exp{\left(-\frac{(\lambda-\lambda_0(1-v_{\text{LOS}}/c))^2}{w^2}\right)}+\text{d}F_{\text{cont}}
\end{equation}
here $\text{d}F_{\text{cont}}=\text{d}V\cdot i_{\text{cont}}/D^2$, and $\text{d}F$ is a function of $h$. Since $h$ is a function of $t$, $\text{d}F$ and $\text{d}f(\lambda)$ are also functions of $t$.

The full stellar disk-integrated CME line profile is an integration of $\text{d}f(\lambda)$ over the total volume
\begin{equation}\label{eq:profile_cme}
    f_{\text{CME}}(t,\lambda)=\int{\frac{\mathrm{d}f(\lambda)}{\mathrm{d} V}} \mathrm{d} V
\end{equation}
here $f_{\text{CME}}$ is time dependent. The unit of $f_{\text{CME}}$ is erg cm$^{-2}$ s$^{-1}$ \AA$^{-1}$. 

\subsubsection{Time-averaged Line Profiles}\label{sec:time-average}
The above processes closely resemble the treatment described in Paper I. From Eq. \ref{eq:profile_cme} we can synthesize the line profile of a stellar CME whose mass center is at a specific height $h$ at a given time $t$. However, as we explained before, in stellar observations, long exposure times are required, so our observed line profiles should be the time-averaged profiles over the duration of the exposure. 

Supposing the exposure time is $\Delta T=t_2-t_1$, which means the exposure is from $t=t_1$ to $t=t_2$, and we have a set of CME line profiles spanning from $t=t_1$ to $t=t_2$, with a temporal cadence of $\text{d}t$. If the cadence is sufficiently small (in our case, 5 s), we can assume that the line profiles remain unchanged during $\text{d}t$ and average the line profiles over the exposure time.

\begin{figure}[htbp!]
\centering
\includegraphics[width=0.8\textwidth]{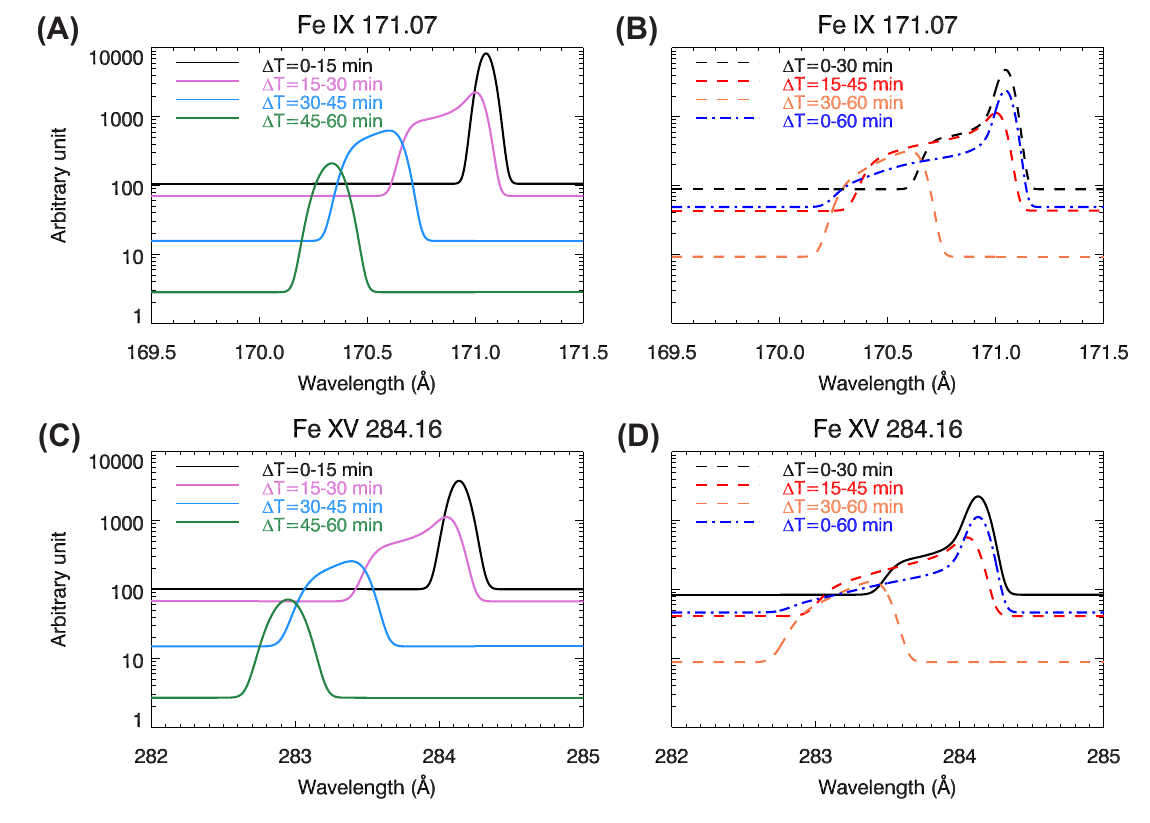}
\caption{Examples of the time-averaged CME line profiles during long exposure. (A-B) Fe \sc{ix}\rm{} 171.07 \AA. The four solid profiles in (A) refer to the exposure time of 15 min with $\Delta T=0-15$ min, $\Delta T=15-30$ min, $\Delta T=30-45$ min and $\Delta T=45-60$ min. The three dashed profiles in (B) refer to the exposure time of 30 min with $\Delta T=0-30$ min, $\Delta T=15-45$ min and $\Delta T=30-60$ min. The dotted-dashed profile in (B) shows the result with an exposure time of $\Delta T=0-60$ min. (C-D) Similar to (A)(B) but for Fe \sc{xv}\rm{} 284.16 \AA.}  \label{fig:time average}
\end{figure}

Figure \ref{fig:time average} shows eight examples of the time-averaged line profiles of Fe \sc{ix}\rm{} 171.07 \AA\ and Fe \sc{xv}\rm{} 284.16 \AA\ with a fixed initial CME area corresponding to 3\% of the stellar disk. The propagation direction considered here is $(\eta=0,\xi=0)$. These profiles show clear distortion from Gaussian shapes, resulting from the rapidly changing CME velocity. During the exposure, the CME travels a certain distance, with its velocity increasing while the emission weakening due to the decrease of the electron density with increased height (see Eq. \ref{eq:iofqh}). Consequently, the line profiles at each time step have different line centroids. As a result, the superposition of these line profiles leads to the distinct shapes depicted in Fig. \ref{fig:time average}. Furthermore, it can be found that when observing from the beginning of the eruption (e.g., from 0 s), the resulted profile will be concentrated in the vicinity of line centroid. This can lead to unresolved blendings of the CME component with the background stationary component, as we will demonstrate in the following sections. It is to be noted that the absolute value of the flux is not important in this work, as we will show in the following that only relative flux is used.

\subsection{Synthesis of Line Profiles during Stellar CMEs}\label{sec:synthesis}

The synthesis of line profiles during stellar CMEs consists of two different steps. First, we need to generate line profiles with varying exposure times and different QS and AR ares. These generated line profiles represent ideal scenarios without taking into account the effects of different noise levels and spectral resolutions. The second step is to add photon noise under different S/N levels to the ideal line profiles, and adjust the spectral resolution to simulate the realistic observed line profiles.

\subsubsection{Stellar line profiles under different exposure times and stellar activities}

The impact of different exposure times has been discussed in the generated time-averaged CME line profiles in Section \ref{sec:time-average}. In our study, we characterized the stellar activity level based on the initial areas (areas at the start of the eruption) of CME region, active regions and quiet Sun regions. To reduce the model complexity, we used a fixed initial CME area (the CME area at the start of the eruption) as $S^0_{\text{CME}}=3\%S_{\star}$ ($S_{\star}$ is the stellar disk area). The AR area $S_{\text{AR}}$ is decomposed into the CME area ($S_{\text{CME}}$) and the area of ARs without CME regions ($S_{\text{AR}^{\prime}}$). $S_{\text{AR}^{\prime}}$ is varying among the values of 2\%, 5\%, 10\% and 20\% of $S_{\star}$, corresponding to $S_{\text{AR}}$=5\%$S_{\star}$, 8\%$S_{\star}$, 13\%$S_{\star}$ and 23\%$S_{\star}$. The QS area is then $S_{\text{QS}}=S_{\star}-S_{\text{AR}}$. Here the CME area refers to its radially projected area onto the stellar disk (not the LOS projected area). From Eq. 4 in Paper I we know that the total flux emitted by a region with area $S$ is $F=(\bar{I}\cdot S)/D^2$. So we have the total flux of AR and QS as $F_{\text{AR}}=(\bar{I}_{\text{AR}}\cdot S_{\text{AR}})/D^2$ and $F_{\text{QS}}=(\bar{I}_{\text{QS}}\cdot S_{\text{QS}})/D^2$ (where $D$ can be simply treated as unity). The corresponding line profiles are 
\begin{equation}
    f_{\text{AR}}(S_{\text{AR}})=\frac{F_{\text{AR}}}{\sqrt{\pi}w_{\text{AR}}}\exp{\left(-\frac{(\lambda-\lambda_0)}{w_{\text{AR}}}\right)}+F_{\text{cont,AR}}
\end{equation}
and
\begin{equation}
    f_{\text{QS}}(S_{\text{QS}})=\frac{F_{\text{QS}}}{\sqrt{\pi}w_{\text{QS}}}\exp{\left(-\frac{(\lambda-\lambda_0)}{w_{\text{QS}}}\right)}+F_{\text{cont,QS}}
\end{equation}
where $w_{\text{AR}}$ is the corresponding 1/e widths of reference line profiles derived from the solar Hinode/EIS observation, and $w_{\text{QS}}$ is assumed to be the same as $w_{\text{AR}}$. The continuum fluxes $F_{\text{cont,AR}}$ and $F_{\text{cont,QS}}$ can be derived based on the continuum in reference line profiles using Eq. 4 in Paper I.

As mentioned in Paper I, the calculation of its total flux and volume of the CME in Sec. \ref{sec:lineprof} and the time-integrated line profiles in Sec. \ref{sec:time-average} already taks into account the changing CME area as it propagates and expands. During the long exposure, we assume that the background ARs and QS regions remain their characteristic emissions (the reference profiles). Therefore, the time-averaged line profiles in ARs and QS regions should be identical to the line profiles without considering time integration effect.

With the time-integrated line profiles of CME regions $f_{\text{CME},\lambda}(t_1, t_2)$ under different exposure times (from $t=t_1$ to $t=t_2$, the total exposure time is $\Delta T=t_2-t_1$) and a fixed initial CME area of 3\% and the varying initial AR areas ($S_{\text{AR}}$), we can construct the stellar line profiles during CME eruptions as follows:
\begin{equation}
    f_{\text{tot}}(t_1, t_2,S_{\text{AR}})=f_{\text{CME},\lambda}(\Delta T)+f_{\text{AR}}(S_{\text{AR}})+f_{\text{QS}}(S_{\text{QS}})
\end{equation}

\subsubsection{Impacts of instrumental parameters}
In addition to the exposure time and stellar activity levels, two instrumental parameters are important for the interpretation of stellar EUV line profiles: S/N and spectral resolution. S/N determines the noise level, which is essential for the detection of CME signals, as they are typically very weak in full disk-integrated EUV line profiles (as discussed in Paper I); spectral resolution is vital for velocity measurement through Doppler effect, as a lower spectral resolution will broaden a line profile and may blend two different components with different velocities. In this part, we will simulate the actual observations by adding noise and varying spectral resolutions for the line profiles. We follow the same methodology to degrade the spectral resolution of the line profiles and add noise with varying S/N as in Paper I. The technical details can be found in Sec. 3.1.2 and Sec. 3.1.3 in Paper I.

\section{Detection of CME Signal and Velocity}

Using the integrals derived in the previous section, we can synthesize various full disk-integrated line profiles during stellar CMEs. 

\begin{figure}[htbp!]
\centering
\includegraphics[width=0.8\textwidth]{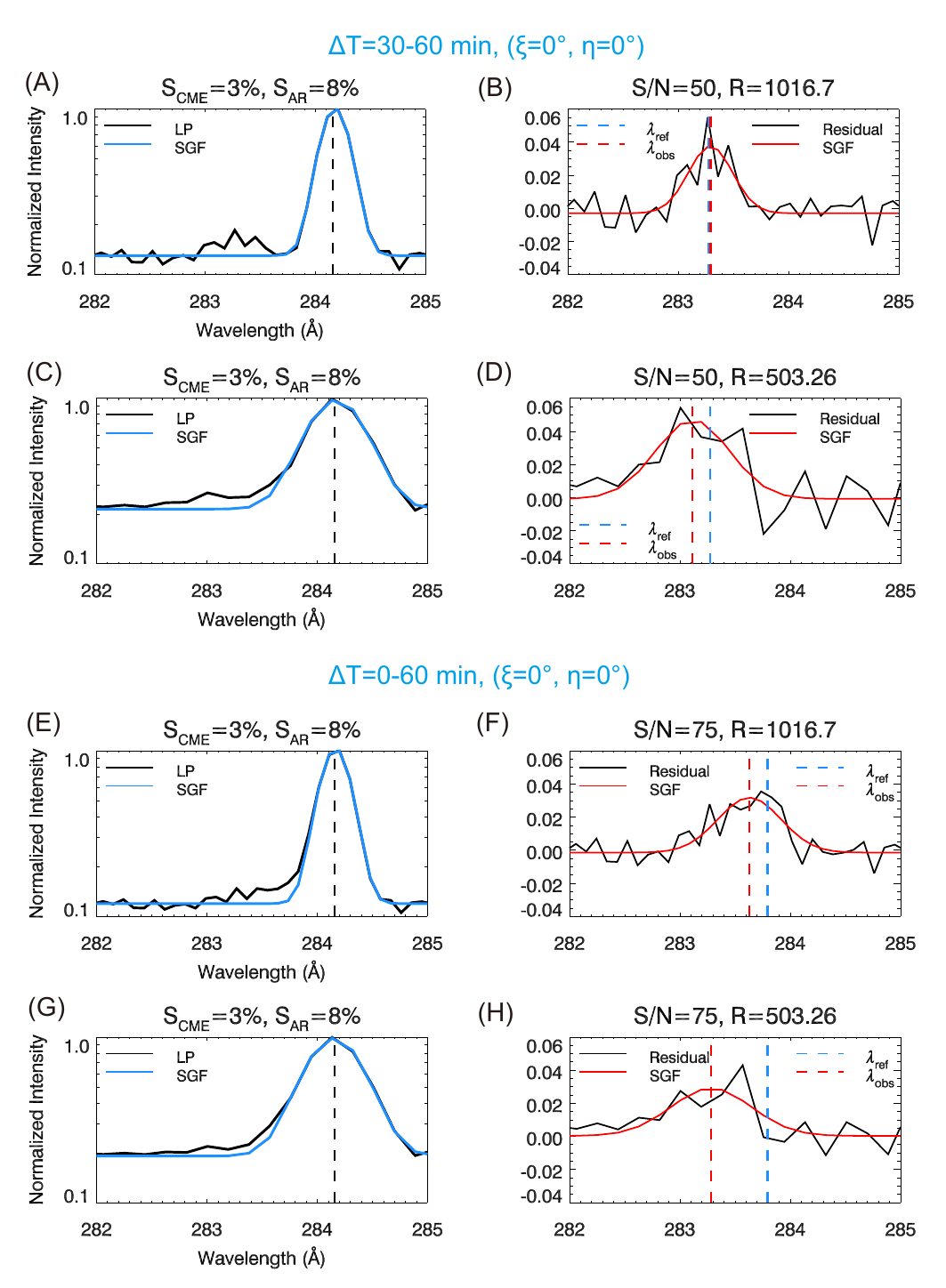}
\caption{The examples of synthesized profiles of Fe \sc{xv}\rm{} 284.16 \AA\ and the procedure of 3$\sigma$ criterion. The left column gives the synthesized line profiles under different instrumental conditions and exposure times (black), and the single-Gaussian fitted profiles for the stationary component contributed by background emission (blue). The black dashed vertical lines mark the rest wavelengths. Note that the spectral pixel size (sampling size) is set to be 1/3 of the instrumental width. The right column shows the residual profiles or the CME component (black solid) and the fitted results (red solid). The active region area is fixed to be 5\% of the full disk area in all cases. The upper four panels are the results under an exposure time of 30 min from $t=30$ min to $t=60$ min and an S/N of 50. The lower four panels are results under $\Delta T=0-60$ min and S/N=75. The first and third rows are results with a spectral resolution of $\delta\lambda\approx 0.28$ \AA\ or a spectral resolving power of $R\approx 1017$. The second and fourth rows are results under a lower spectral resolution ($\delta\lambda\approx 0.56$ \AA\ or $R\approx 503$. In the right column, the vertical blue dashed line marks the reference centroid wavelength (velocity) of the CME component (see main text for details). The vertical red dashed line marks the fitted centroid wavelength from the residual profiles. It can be found that with a lower spectral resolution, the deviation of the fitted centroid from the reference one is larger. If we integrate the observations from the beginning of the eruption rather than focus on a later phase, in this case, the deviation is also larger.}  \label{fig:stellar spectra}
\end{figure}

When interpreting the EUV line profiles from stellar CME observations, the most important objective is to identify the signals contributed by stellar CMEs. In our study, for CMEs traveling towards the observers, their LOS components should result in blue-wing enhancements or a secondary blue-shifted component in the line profile. To evaluate whether we can observe CME signals from the synthesized line profiles, we adopted the same $3\sigma$ criterion as in Paper I: a spectral line profile with blue-wing enhancement can be separated into a stationary component contributed by the background emission and a blue-shifted CME component; if the peak intensity of the CME component is larger than three times the standard deviation of the background continuum, we define the line profile as ``CME-detectable". For ``CME-detectable" line profiles, we can do a single-Gaussian fit to the CME component (or the residual component, obtained by subtracting fitted single-Gaussian from the original profile) to derive its velocity. A further comparison between the calculated velocity and a reference velocity will give us more information on whether we can diagnose the CME velocity with sufficient accuracy. Figure \ref{fig:stellar spectra} shows examples of different stellar Fe \sc{xv}\rm{} 284.16 \AA\ spectra under different instrumental configurations and exposure times. The left column shows the synthesized line profiles (black) and single-Gaussian fitted profiles of the stationary components. The right column shows the residual profiles (black solid profiles) and corresponding single-Gaussian fittings (red solid profiles). The vertical blue and red dashed lines mark the reference blue-shift and the fitted blue-shift from residual profiles, respectively. Here the reference blue-shift is the blue-shift derived under the optimal instrumental conditions in this study (S/N$=250,\ R=R_{\text{EIS}}\approx 3936$).

\begin{figure}[htbp!]
\centering
\includegraphics[width=0.95\textwidth]{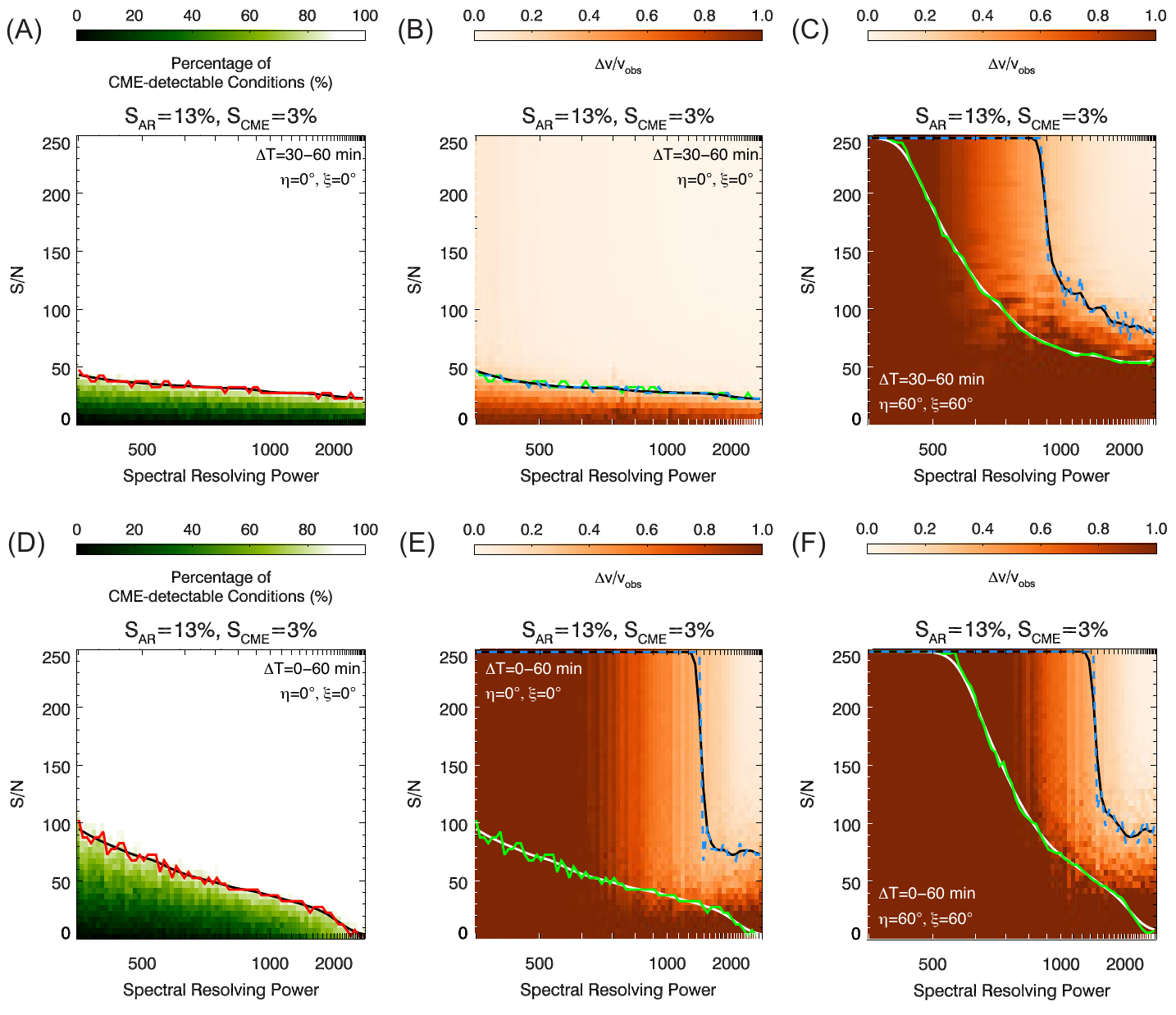}
\caption{Examples on the derivation of suitable instrumental parameters to detect CME signals and velocities using Fe \sc{xv}\rm{} 284.16 \AA. The top panels are results under $\Delta T=30-60$ min and the lower panels are results under $\Delta T=0-60$ min. The left column shows two maps showing the derivation of ``CME-detectable" conditions only. The red and black curves are the critical curves representing the minimum required instrumental configurations and the Gaussian-smoothed curves, respectively. The other two columns are similar to the left column but included the accuracy of derived CME velocities. The middle column shows results for CMEs erupting along the star-Earth direction, while the right column provides results with a large propagation direction of $(\eta=60^{\circ},\ \xi=60^{\circ})$. The AR area is fixed to be 10\% for all cases. Note that the X-axis is not uniform, as we converted the uniform spectral resolution $\delta\lambda$ to spectral resolving power $R$.}  \label{fig:sigma vel map}
\end{figure}

We can perform the above procedures for each line profile generated under different exposure times ($\Delta T$), instrumental configurations ($\delta\lambda_{\text{instr}}$ and S/N) and different stellar activity levels ($S_{\text{AR}}$ and $S_{\text{QS}}$). Similar to the procedures described in Paper I, for each set of variables, we generated 200 different line profiles and conducted the $3\sigma$ test. If 80\% of the line profiles are classified as ``CME-detectable", the instrumental configuration used during their synthesis is defined as the critical instrumental parameters required for detecting stellar CME signals using line profile asymmetry method. The frontier of these critical instrumental parameters forms a critical curve. Figure \ref{fig:sigma vel map}(A) and (D) give two examples of the critical curves. The x-axis is the the spectral resolving power $R=\lambda_0/\delta\lambda$, where $\lambda_0$ is the line centroid wavelength and $\delta\lambda$ is the spectral resolution, and the y-axis is the signal-to-noise ratio (S/N). Each pixel in the map represents the number of ``CME-detectable" cases out of the 200 different profiles. The red curve is the original critical instrumental parameter curve, indicating that for each pair of ($\delta\lambda$ , S/N) above this curve, we have a chance of over 80\% to detect the CME signal from the corresponding line profiles. The black curve is a Gaussian-smoothed result of the red one. 

In addition to the detection of CME signals, another crucial objective is the accurate derivation of the CME velocity. This parameter is obtained by applying a single-Gaussian fit to the residual profiles (as shown in examples in Fig. \ref{fig:stellar spectra}). Although it has been shown that the long exposure times can significantly distort the CME profiles from Gaussian shapes, we adhere to the single-Gaussian fitting or the method of moments (which yield similar results) due to the difficulty of separating different components (e.g., multi-Gaussian components) in observed stellar profiles without prior knowledge of these components. This is particularly relevant in real observations, as the CME may contain regions with different velocities, resulting in peculiar profiles. To assess the accuracy of the derived velocity, we first need to choose a reference velocity. For Sun-as-a-star observations where long exposure times are not necessary, the reference velocity can be simply selected as the velocity at the corresponding time. In contrast, stellar observations require long exposures, making the selection of a reference velocity un-intuitive. In this work, we choose the velocity derived under the optimal instrumental conditions (S/N$=250,\ R=R_{\text{EIS}}$) as the reference velocity ($v_{\text{ref}}$). Similar to the previous paragraph, for each set of variables, we have 200 different line profiles and corresponding residual profiles. If a profile meets the $3\sigma$ criterion, we perform a single-Gaussian fit to the residual profile and obtain a velocity value ($v_{\text{obs}}$). If the criterion is not met, $v_{\text{obs}}$ is set to $2v_{\text{ref}}$ to ensure the deviation is set to 100\%. We then calculate the average value of the differences between the 200 different $v_{\text{obs}}$ and $v_{\text{ref}}$, as well as their relative values with respect to the reference velocity ($\Delta v/v_{\text{ref}}$, where $\Delta v=|v_{\text{obs}}-v_{\text{ref}}|$). 

The middle and right columns of Figure \ref{fig:sigma vel map} provide examples of the velocity difference maps. Each pixel in the maps represents the average value of $\Delta v/v_{\text{ref}}$, ranging between 0 and 1. If, for a pair of instrumental parameters, $\Delta v/v_{\text{ref}}$ is smaller than 0.3 (the uncertainty of the derived velocity is within 30\% of the reference value), the instrumental parameters are deemed capable of accurate velocity measurement. This also yields a set of critical instrumental parameters and the corresponding critical curve for accurate velocity diagnostics. For each spectral resolution $\delta\lambda$, the black curves in Fig. \ref{fig:sigma vel map}(B)(C)(E)(F) represent the critical curves, while the blue dashed curves are Gaussian-smoothed versions. The green and white curves in these four panels correspond to the curves solely focused on detection of CME signals from Fig. \ref{fig:sigma vel map}(A) and (D).

The top panels in Fig. \ref{fig:sigma vel map} are results for an exposure time of $\Delta T=30-60$ min, while the bottom panels are for $\Delta T=0-60$ min. The middle column shows results for propagation directions of $(\eta=0^{\circ},\ \xi=0^{\circ})$, while the propagation directions in the right column are $(\eta=60^{\circ},\ \xi=60^{\circ})$. The active region area is fixed to be 10\% for all cases.

\begin{figure}[htbp!]
\centering
\includegraphics[width=0.8\textwidth]{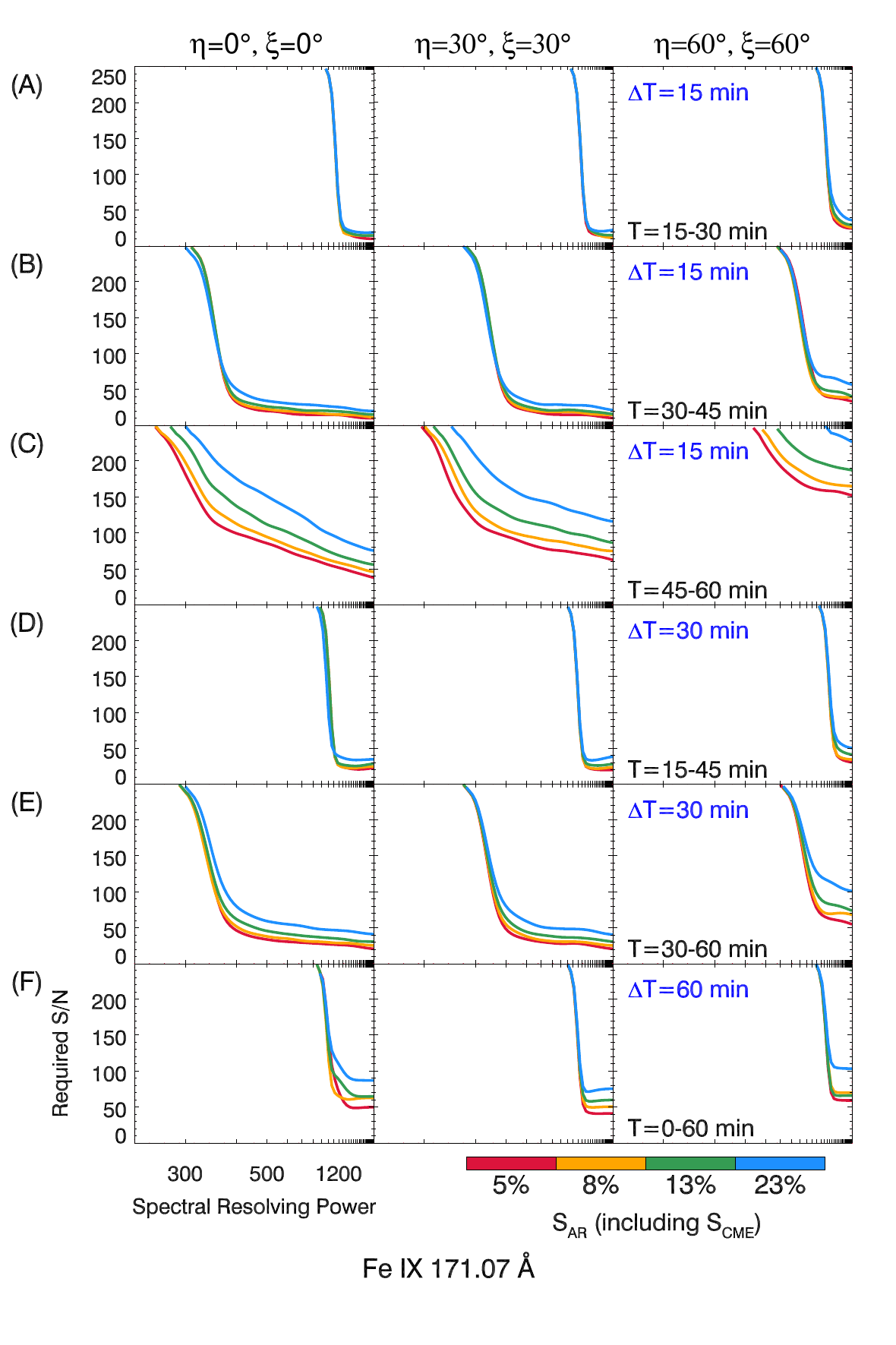}
\caption{The overall constraints on instrumental configurations for different conditions. The x-axis is the spectral resolving power which is determined by the spectral resolution, the y-axis is the signal-to-noise ratio. Each row refers to different exposure times, each column refers to different propagation directions. The initial CME area is fixed to be 3\% of the full disk. Different colored curves refer to the critical instrumental configurations under different AR areas to detect CME signals and determine their velocities within 30\% accuracy. The regions in the graph plane that are above the curves are good for the detection of stellar CMEs, while those regions below the curves are bad for detectability.}  \label{fig:detect171}
\end{figure}

\begin{figure}[htbp!]
\centering
\includegraphics[width=0.8\textwidth]{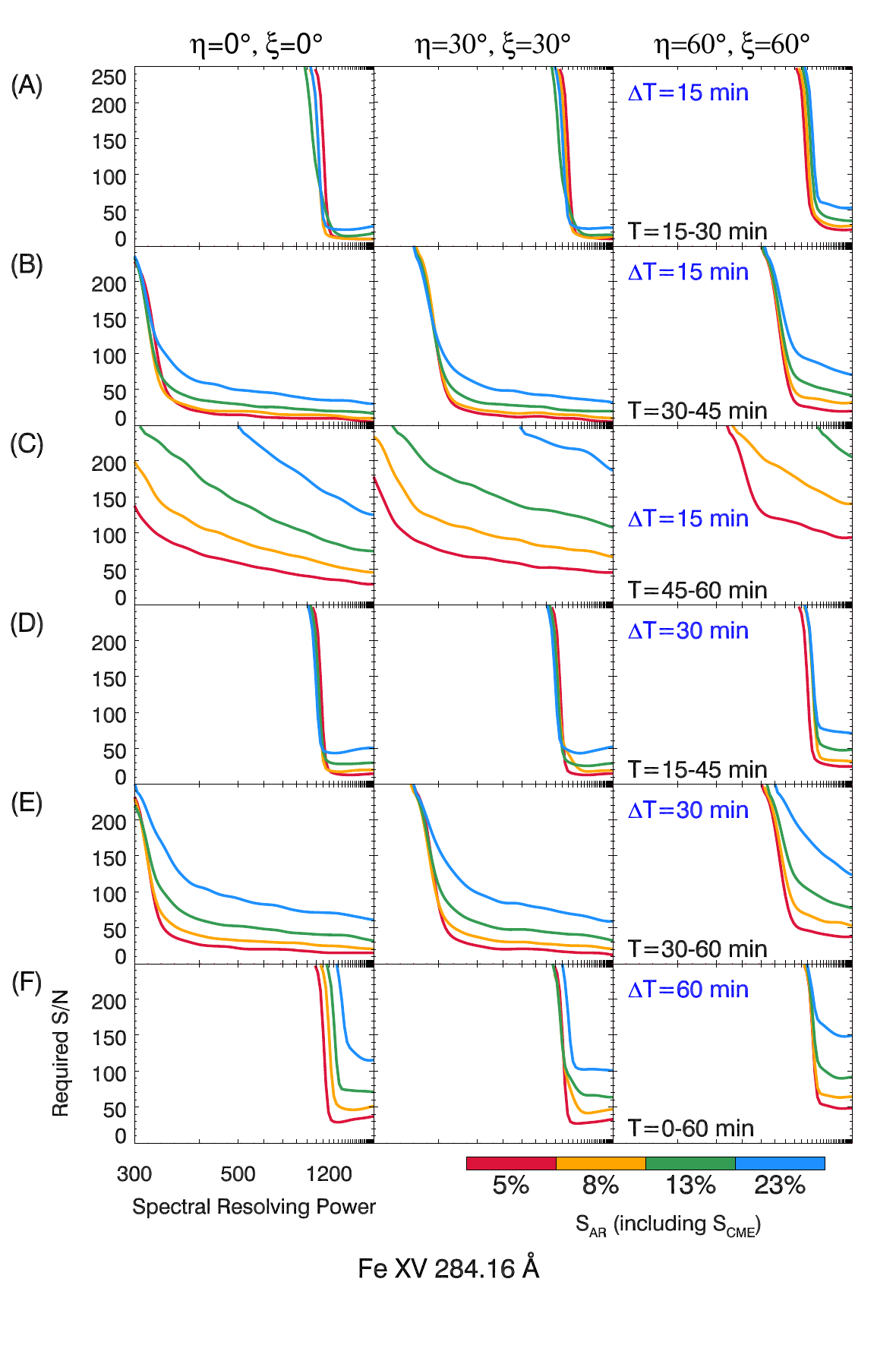}
\caption{Similar to Fig. \ref{fig:detect171} but for the hotter Fe \sc{xv}\rm{} 284.16 \AA.}  \label{fig:detect284}
\end{figure}

Following the aforementioned procedure, we can determine the constraints on instrumental configurations under various conditions for both Fe \sc{ix}\rm{} 171.07 \AA\ and Fe \sc{xv}\rm{} 284.16 \AA. The results are shown in Figures \ref{fig:detect171} and \ref{fig:detect284}, respectively. In each diagram, the x- and y-coordinates are spectral resolution and signal-to-noise ratio, respectively. Each row and column correspond to different exposure times ($\Delta T$) and propagation directions $(\eta,\ \xi)$, respectively. The curves of different colors are the overall critical instrumental configuration curves under different AR areas. These critical curves are results that ensure both accurate detection of CME signals and precise measurements of CME velocities.


\section{Results and Discussions}
\subsection{The results}

Starting from the solar EUV spectral line profiles during a mass ejection observed by Hinode/EIS, and assuming a threefold increase in the electron density for the CME region compared to its background region, we constructed the reference line profiles for CME, AR and QS regions at the initial erupting height. Using a kinematic model and an analytical geometric model, we synthesized full disk-integrated line profiles of Fe \sc{ix}\rm{} 171.07 \AA\ and Fe \sc{xv}\rm{} 284.16 \AA\, considering various stellar activity levels (areas of AR and QS regions), exposure times, CME propagation directions, and instrumental configurations (signal-to-noise ratio and spectral resolution). 

To accurately detect stellar CME velocities, we focused on the blue-shifted emissions contributed by the erupting CME, which lead a blue-wing enhancement in the line profile. By applying the 3-sigma criterion and conducting Monte-Carlo simulations, we determined the overall instrumental constraints. The results for both spectral lines are shown in Figs. \ref{fig:detect171} and \ref{fig:detect284}. 

The results highlight the role of different exposure times and observations during different CME propagating phases for successful detection of stellar CMEs through EUV spectral line profiles. If we observe from, or even before, the initiation of the CME eruption, a high spectral resolving power or spectral resolution is required (e.g., $R>1200$, see rows (A)(D)(F) of Figs. \ref{fig:detect171} and \ref{fig:detect284}). This is because although at the initial propagating phase of a stellar CME, the CME is situated at a relatively low height, resulting in higher density and subsequently stronger emission, the blue-shift is minimal due to the CME's limited velocity at the onset of the eruption. As a consequence, the peaks of the CME line profiles are close to the rest wavelength, where the background component dominates. As the CME propagates, its height increases, emission decreases and velocity becomes larger. When integrating emissions from the beginning of the eruption, or even before it starts, the line profiles are dominated by emissions from lower heights with smaller velocities (as can be seen in the time-averaged CME line profiles during $\Delta T=0-15\ \text{min},\ \Delta T=0-30\ \text{min}$ and $\Delta T=0-60\ $min in Figure \ref{fig:time average}). To discern the line profiles contributed by CMEs from the background emissions, a high spectral resolution is necessary to resolve the subtle differences. However, if the observation begins at a later time during the propagation, since the CME velocity/blue-shift is large, the blue-shifted component is distinct from the line centroid or the background stationary component. In this case, a very high spectral resolution is not required to resolve the CME components. Nevertheless, this is highly dependent on the propagation directions of the stellar CME. From rows (C)(E) of Figures \ref{fig:detect171} and \ref{fig:detect284}, it is obvious that with a tilted propagation direction away from the LOS direction, the LOS projected velocity will decrease, leading to a smaller blue-shift. Hence, a higher spectral resolution is required to observe CMEs where the angle between the propagation direction and the LOS is larger. 

In our investigation, we also explored various combinations of propagation directions $(\eta,\xi)$ to assess their effects. We consistently observed that as the values of $\eta$ and $\xi$ increase, the LOS velocity will decrease. Consequently, a higher spectral resolution is necessary to accurately detect these smaller LOS velocities.

Furthermore, the stellar activity levels also affect the instrumental requirements. For active stars with larger AR areas, the background AR emissions are stronger, so higher spectral resolution and signal-to-noise ratio are required. 

{It is also worth noting that we choose a threefold increase in the density value for the CME regions compared to the original solar CME observation. However, this $3\times$ factor can be changed as different CMEs may possess different properties. Since our detection method is based on the identification of CME component from observed line profiles, the detectability is highly dependent on the intensity of the CME component. With a larger reference CME density, the CME component from the synthesized line profiles will be more discernible, thus a lower S/N and spectral resolution can be used to detect such signals. Conversely, if the reference CME density is lower than the current threefold increased value, more strict instrumental configurations will be required. Nevertheless, based on the conventional choice of CME regions in MHD modeling \citep[e.g.,][]{2019ApJ...884L..13A}, such a threefold increase is still reasonable.}

To summarize the two figures, for active regions covering 5-8\% of the stellar disk, using Fe \sc{ix}\rm{} 171 \AA\ and Fe \sc{xv}\rm{} 284 \AA, if we observe a later stage of CME propagation, and if the CME propagates along the Earth-star direction, a spectral resolving power of at least 300-400 ($\delta\lambda\approx 0.5-0.6$ \AA\ for Fe \sc{ix}\rm{} 171 \AA\ and $\delta\lambda\approx 0.7-0.8$ \AA\ for Fe \sc{xv}\rm{} 284 \AA) and an S/N of at least 30-40 are required for accurate detection the stellar CME velocity. For $R\approx 500$, the required S/N can be reduced to around 20-30. In the case of tilted eruptions, a higher spectral resolving power is required. If the observation starts at or before the eruption, a substantially higher spectral resolving power of at least 1200 ($\delta\lambda\approx 0.15$ \AA\ for Fe \sc{ix}\rm{} 171 \AA\ and $\delta\lambda\approx 0.2$ \AA\ for Fe \sc{xv}\rm{} 284 \AA) and an S/N of at least 50 are necessary. In summary, for stellar EUV spectral observations with exposure times longer than 15 min, with a moderate spectral resolving power and signal-to-noise ratio, it is possible to detect the stellar CMEs erupting towards the observer. To detect CMEs with differently tilted propagation directions, a high spectral resolution will be required.

\subsection{The impacts of interstellar absorption}

The impacts of assumptions and models, as well as the effects of adjacent spectral lines and flare signals contamination, have been discussed in Paper I. However, when conducting stellar EUV observations, an additional important concern arises: the impact of strong absorption from the interstellar medium (ISM) within EUV passband. In the EUV wavelength range, the main sources of opacity are interstellar hydrogen and helium \citep[e.g.,][]{1994AJ....107.2108R,2019LNP...955.....L}. Neutral hydrogen is the major component in ISM, and the bound-free transition of neutral hydrogen contributes significantly to the opacity for wavelength shorter than Lyman edge \citep[$\sim 300-912$ \AA,][]{1994AJ....107.2108R}. The absorption of EUV emissions is influenced by the column density of interstellar hydrogen, as shown in Figure 6.14 from \cite{2019LNP...955.....L}, which provides examples of optical depth for some nearby stars. In the worst cases, EUV emissions with wavelengths longer than roughly 360 \AA\ are almost completely absorbed due to the optical depth larger than unity \citep{1994AJ....107.2108R,2005ApJ...622..680R}. Therefore, observing the full EUV spectrum, even for the nearest stars, becomes almost impossible.  

For wavelengths between approximately 30 \AA\ and 300\AA, neutral helium contributes to the majority of opacity \citep{1994AJ....107.2108R}. For nearby stars (e.g., within 30 pc), the absorption in this wavelength range is only partial and remains relatively constant within a narrow wavelength range in the surrounding of the spectral lines used in our work. This is the reason why we choose specifically EUV spectral lines with wavelengths shorter than 360 \AA\ for investigation. For the two spectral lines we select, the interstellar absorption can be treated as a constant partial absorption to both line intensity and continuum. In our analysis, we only use the relative line profiles, with their maximum scaled to unity, which minimizes the impact of the constant absorption on the profiles. However, this constant partial absorption does result in a reduced number of detected photons and consequently a lower S/N. In cases where we observe nearby stars with insignificant absorption, the results will not be significantly altered. 

\begin{figure}[htbp!]
\centering
\includegraphics[width=0.8\textwidth]{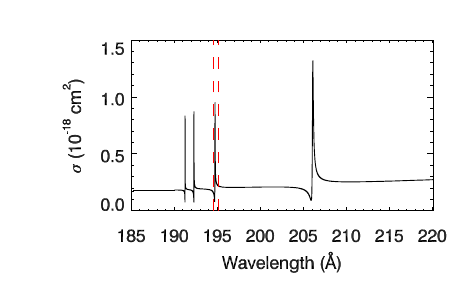}
\caption{The absorption cross section as a function of wavelength in the range of 185 \AA\ to 220 \AA. The interstellar hydrogen and helium column densities are set to be $10^{18}\ \text{cm}^{-2}$ and $10^{17}\ \text{cm}^{-2}$, respectively. The four asymmetric absorption profiles can be found in this wavelength range. The vertical red dashed lines mark the wavelength positions of 194.6 \AA \ and 195.2 \AA.}  \label{fig:ismeuv}
\end{figure}

Nevertheless, it is worth noting that we do not use the Fe \sc{xii}\rm{} 195.12 \AA\ line as we did in Paper I, despite it being one of the strongest emission lines in the stellar coronae. This is due to a very pronounced asymmetric absorption near this wavelength. When a photon is absorbed by neutral helium, one of the helium's electron can be directly ionized while the other remains in a bound state. This process is known as direct photonionization. Alternatively, the two electrons of neutral helium can be first excited to excited states (doubly excited states), and then one electron decays to a lower energy level, releasing energy that ionizes the other electron. This second process, which does not involve photon absorption, is called autoionization \citep{2015aas..book.....P}. The autoionizing transition of helium shows strong resonance structures, leading to the four different asymmetric absorption profiles between 170 \AA\ and 210 \AA\ \citep{1994AJ....107.2108R}. Figure \ref{fig:ismeuv} shows the absorption cross section between 185 \AA\ and 220 \AA\ with interstellar hydrogen and helium column densities of $10^{18}\ \text{cm}^{-2}$ and $10^{17}\ \text{cm}^{-2}$, respectively. We use the IDL function \it{ismeuv.pro}\rm{} to calculate the cross section. It is evident that in the vicinity of Fe \sc{xii}\rm{} 195.12 \AA\ line, the absorption cross section exhibits a typical asymmetric Fano profile \citep[e.g.,][]{1961PhRv..124.1866F}. This strong asymmetric absorption significantly affects the observed Fe \sc{xii}\rm{} 195.12 \AA\ profile, making it inappropriate for the detection of stellar CMEs. While it is possible to reconstruct the absorption cross section and recover the ``unabsorbed" line profiles, this process heavily relies on accurate information regarding the column densities of interstellar hydrogen and helium. Therefore, we do not include this line in our analysis. In contrast, the two EUV lines used in this work are not affected by these asymmetric absorptions.

\section{Summary}

Coronal mass ejections are the most important sources of space weather effects for both planets in the solar system and exoplanets in other stellar systems. They also play a key role in the habitability of exoplanets. Therefore, the detection of stellar CMEs is important for the study of extraterrestrial space weather and the habitability of exoplanets. Given that numerous strong EUV lines are formed in the coronae of solar-type stars, they may be used for the detection of coronal material in stellar CMEs. The blue-shifts caused by propagating CMEs can result in a blue-wing enhancement in the spectral line profile. Based on the analysis to the spectral line profile asymmetry, it becomes possible to identify the CME signals and measure their velocities. However, due to the strong interstellar absorption in EUV band and the lack of EUV spectral observations for solar-type stars, it is unclear whether we can detect stellar CMEs or under what conditions can we detect them. In this work, based on solar EUV spectral observations, we constructed a geometric model for stellar CMEs, and synthesized the full disk-integrated line profiles of Fe \sc{ix}\rm{} 171.07 \AA\ and Fe \sc{xv}\rm{} 284.16 \AA. We also considered the effects of different propagation directions and different exposure times. 

Our findings indicate that the accurate detection of stellar CME signals and velocities is highly dependent on the observed CME propagation phases and directions. Under typical solar-type star conditions, if we observe a later phase of an erupting CME with a small tilt angle away from LOS, a spectral resolving power of at least 300-400 and an S/N of at least 30-40 (or $R\approx 500$ and S/N$\approx$20-30) are required to accurately detect the stellar CME velocity. If we observe the CME from its initiation or if the CME propagates at large angles away from the LOS, more stringent conditions are required. In such cases, we need a much higher spectral resolving power of at least 1200 and an S/N of at least 50. 

In summary, it is feasible to detect CMEs on solar-type stars erupting towards the observer using an EUV spectrometer with moderate spectral resolving power and S/N. Since CMEs erupt along various directions, long-time observations with such an instrument on multiple featured targets in the neighborhood of solar system are required to achieve more successful detections.

\begin{acknowledgments}
This research is supported by the NSFC grant 12250006, National Key R\&D Program of China No. 2021YFA0718600，the Frontier Scientific Research Program of Deep Space Exploration Laboratory (2022-QYKYJH-ZYTS-016) and the New Cornerstone Science Foundation through the XPLORER PRIZE. We are grateful to Dr. Yajie Chen (MPS) and Mr. Xianyu Liu (PKU/UMich) for their helpful discussions.

\end{acknowledgments}

\appendix

\section{The synthesis of reference line profiles}\label{app:appendixa}

In Section \ref{sec:refline}, we establish a reasonable assumption that the electron density in CME regions can be three times higher than that in the background ARs. From \cite{2012ApJ...748..106T}, it is estimated that the electron density in the erupting loop regions is $10^{10}\ \text{cm}^{-3}$, similar to the electron density in the surrounding active regions. Therefore, we simply choose a threefold increase in the original observed CME density. Additionally, we demonstrate that the EUV line intensity is not solely dependent on the square of electron density but also on another function, denoted as $Q(n_e)$. 

Based on the observed Fe \sc{xii}\rm{} 195.12 \AA\ and Fe \sc{xv}\rm{} 284.16 \AA\ profiles in Fig. 1 of Paper I, we can derive the fitted CME profile as $f_0(\lambda)=\dfrac{I}{\sqrt{\pi}w}\exp{-\dfrac{(\lambda-\lambda_0)^2}{w^2}}+I_B$ ($\lambda_0$ is the centroid wavelength, $I$ is the total line intensity and $I_B$ is the continuum). Referring to Eq. \ref{eq:qfunc}, we can calculate the new CME line intensity considering a three-fold increase in the electron density:
\begin{equation}
    I_{3\times}=\frac{Q(3n_e)}{Q(n_e)}(\frac{3n_e}{n_e})^2\cdot I=9I\cdot\frac{Q(3n_e)}{Q(n_e)}
\end{equation}
The new CME line profile is $f_{3\times}(\lambda)=\dfrac{I_{3\times}}{\sqrt{\pi}w}\exp{-\dfrac{(\lambda-\lambda_0)^2}{w^2}}+I_B$. 

For Fe \sc{ix}\rm{} 171.07 \AA, since it was not directly observed in the EIS observation we used, we need to estimate its intensity assuming the same physical conditions as Fe \sc{xii}\rm{} 195.12 \AA\ within the same regions. Using CHIANTI calculations, we find that for typical active regions, the intensity ratio between Fe \sc{ix}\rm{} 171.07 \AA\ and Fe \sc{xii}\rm{} 195.12 \AA\ is $\sim 1.88$. Thus, we have $I_{\text{AR},171}=1.88 I_{\text{AR},195}$, where $I_{\text{AR},195}$ is a known value obtained from observation. We then calculated the line width of Fe \sc{ix}\rm{} 171.07 \AA\ ($w_{171}$) by convolving its thermal width (determined by the ion temperature and ion mass), the non-thermal width \citep[assumed to be $\sim 30\ \text{km}\ \text{s}^{-1}$, e.g.,][]{2016ApJ...820...63B} and the EIS instrumental width. The continuum in the wavelength near 171 \AA\ is around 0.85 of that near 195 \AA\ according to CHIANTI calculation. The line profile in AR is then $f_{\text{AR},171}(\lambda)=\dfrac{I_{\text{AR},171}}{\sqrt{\pi}w_{171}}\exp{-\dfrac{(\lambda-\lambda_0)^2}{w_{171}^2}}+I_{B,171}$. From \cite{2007PASJ...59S.857Y} and CHIANTI calculations we also have the typical Fe \sc{ix}\rm{} 171.07 \AA\ intensity ratio between AR and QS to be $\sim 7$, allowing us to construct line profiles in QS regions. Here the line width in QS regions is assumed to be the same as in AR without eruptions, as the line width observed by Hinode/EIS is dominated by instrumental broadening. Following the same procedure described in the previous paragraph, the CME line profile can also be derived.

\bibliography{manuscript}{}
\bibliographystyle{aasjournal}

\end{CJK*}
\end{document}